\documentclass[aps,showkeys,showpacs,twocolumn,showpacs,preprintnumbers,amsmath,amssymb,superscriptaddress,floatfix,nofootinbib]{revtex4}
\usepackage{graphicx,color,dcolumn,booktabs,bm}
\usepackage{longtable,lscape}
\usepackage{txfonts}
\usepackage{overpic}
\usepackage{epsfig}
\usepackage{amssymb}
\usepackage{rotating}
\usepackage{epstopdf}
\usepackage{indentfirst}
\usepackage{feynmf}   %{feynmp}
\usepackage{slashed}  %for Feynman symbols
\usepackage{cases}
\usepackage{color}
\usepackage{multirow}
\usepackage{graphicx,color,dcolumn,booktabs,bm}
\usepackage{cases}
\usepackage{array}

\graphicspath{{Figures/}} %
\usepackage[colorlinks, citecolor=blue,anchorcolor=red,menucolor=red, linkcolor=red,filecolor=red,runcolor=red,urlcolor=blue,frenchlinks=red]{hyperref}
\begin{document}

\title{Canonical interpretations of the newly observed $\Xi_c(2923)^0$, $\Xi_c(2939)^0$ and $\Xi_c(2965)^0$ resonances}

\author{Qi-Fang L\"{u}} \email[]{lvqifang@hunnu.edu.cn}
\affiliation{  Department
of Physics, Hunan Normal University,  Changsha 410081, China }

\affiliation{ Synergetic Innovation
Center for Quantum Effects and Applications (SICQEA), Changsha 410081,China}

\affiliation{  Key Laboratory of
Low-Dimensional Quantum Structures and Quantum Control of Ministry
of Education, Changsha 410081, China}

\begin{abstract}

Three neutral resonances $\Xi_c(2923)^0$, $\Xi_c(2939)^0$ and $\Xi_c(2965)^0$ have been observed in the $\Lambda_c^+ K^-$ mass spectrum by the LHCb Collaboration. Given the $\Xi_c$ and $\Xi_c^\prime$ mass spectra predicted by the constituent quark models, these three resonances are tentatively treated as the $\lambda-$mode $\Xi_c(2S)$, $\Xi_c^\prime(2S)$, $\Xi_c^{\prime *}(2S)$, and $\Xi_c^\prime(1P)$ states, and their strong decay behaviors are calculated within the $^3P_0$ model.  Our results indicate that the $\Xi_c(2923)^0$ and $\Xi_c(2939)^0$ can be clarified into the $J^P=3/2^-$ and $5/2^-$ $\Xi_c^\prime(1P)$ states respectively, and the $\Xi_c(2965)^0$ can be regarded as the $J^P=1/2^+$ $\Xi_c^\prime(2S)$ state. Also, we suggest that the previous observed $\Xi_c(2930)$ may be the overlap of two structures $\Xi_c(2923)$ and $\Xi_c(2939)$, and the $\Xi_c(2970)$ should be the same state as $\Xi_c(2965)$. Other theoretical information on the missing $\Xi_c(2S)$, $\Xi_c^{\prime *}(2S)$, and $\Xi_c^\prime(1P)$ states may be helpful for future experimental searches.
\end{abstract}


\maketitle

\section{Introduction}{\label{introduction}}

A charmed strange baryon $\Xi_c$ is composed of a heavy charm quark, a light up or down quark, and a strange quark. Under the classification of constituent quark model, the $\Xi_c$ baryons can be divided into two families, the antisymmetric flavor configuration $\bar 3_F$ and symmetric flavor configuration $6_F$. To distinguish these two families, the states with $\bar 3_F$ flavor part are denoted as $\Xi_c$ states, and the notation $\Xi_c^\prime$ stands for the states belonging to the $6_F$ flavor configuration. Unlike the $\Lambda_c$ and $\Sigma_c$ states with different isospins, both $\Xi_c$ and $\Xi_c^\prime$ baryons have the same isospin $I=1/2$. Since the observed resonances may belong to the flavor $\bar 3_F$ or $6_F$ and can not be distinguished through the quantum numbers, it is more complicated and challenging to study these states both experimentally and theoretically.

Very recently, the LHCb Collaboration reported three new resonances $\Xi_c(2923)^0$, $\Xi_c(2939)^0$ and $\Xi_c(2965)^0$ in the $\Lambda_c^+ K^-$ mass spectrum~\cite{Aaij:2020yyt}. The large significance indicates that these three baryons are unambiguously observed, and the measured masses and total widths are presented as follows,
\begin{eqnarray}
m[\Xi_c(2923)^0] = 2923.04\pm0.25\pm0.20\pm0.14~\rm{MeV},
\end{eqnarray}
\begin{eqnarray}
\Gamma[\Xi_c(2923)^0] = 7.1\pm0.8\pm1.8~\rm{MeV},
\end{eqnarray}
\begin{eqnarray}
m[\Xi_c(2939)^0] = 2938.55\pm0.21\pm0.17\pm0.14~\rm{MeV},
\end{eqnarray}
\begin{eqnarray}
\Gamma[\Xi_c(2939)^0] = 10.2\pm0.8\pm1.1~\rm{MeV},
\end{eqnarray}
\begin{eqnarray}
m[\Xi_c(2965)^0] = 2964.88\pm0.26\pm0.14\pm0.14~\rm{MeV},
\end{eqnarray}
\begin{eqnarray}
\Gamma[\Xi_c(2965)^0] = 14.1\pm0.9\pm1.3~\rm{MeV}.
\end{eqnarray}
The LHCb Collaboration also pointed out several equalities of mass gaps,
\begin{eqnarray}
m[\Omega_c(3050)^0] - m[\Xi_c(2923)^0] \simeq 125~\rm{MeV},
\end{eqnarray}
\begin{eqnarray}
m[\Omega_c(3065)^0] - m[\Xi_c(2939)^0] \simeq 125~\rm{MeV},
\end{eqnarray}
\begin{eqnarray}
m[\Omega_c(3090)^0] - m[\Xi_c(2965)^0] \simeq 125~\rm{MeV},
\end{eqnarray}
\begin{eqnarray}
m[\Xi_c(2923)^0] - m[\Sigma_c(2800)^0] \simeq 125~\rm{MeV},
\end{eqnarray}
which strongly suggests that the $\Xi_c(2923)^0$, $\Xi_c(2939)^0$ and $\Xi_c(2965)^0$ should be the corresponding charmed strange partners of the $\Omega_c(3050)^0$, $\Omega_c(3065)^0$, and $\Omega_c(3090)^0$, respectively. Also, the $\Sigma_c(2800)^0$ may be the non-strange partner of the $\Xi_c(2923)^0$ and $\Omega_c(3050)^0$. A recent work with QCD sum rule suggests that these three newly observed states can be explained as $P-$wave $\Xi_c^\prime$ baryons~\cite{Yang:2020zjl}.

From the Review of Particle Physics~\cite{Tanabashi:2018oca}, there exist ten observed $\Xi_c$ or $\Xi_c^\prime$ baryons. Plenty of theoretical works have been done to investigate their inner structures~\cite{Ebert:2007nw,Roberts:2007ni,Ebert:2011kk,Chen:2014nyo,Chen:2016iyi,Chen:2007xf,Liu:2012sj,Mu:2014iaa,Ye:2017yvl,Ye:2017dra,Wang:2017kfr,
Valcarce:2008dr,Chen:2015kpa,Cheng:2006dk,Cheng:2015naa,Cheng:2017ove,Aliev:2018ube,Shah:2016nxi,Jia:2019bkr,JimenezTejero:2009vq,Lu:2014ina,Yu:2018yxl,Nieves:2019jhp,
Zhao:2016qmh,Yao:2018jmc,Liu:2007ge,Chen:2017aqm,Guo:2008he,Chen:2016phw,Zhang:2008pm,Wang:2010it,Kawakami:2019hpp}. Three lowest structures correspond to the ground states $\Xi_c$, $\Xi_c^\prime$, and $\Xi_c^{\prime *}$ undoubtedly. The $\Xi_c(2790)$ and $\Xi_c(2815)$ should belong to the $\Xi_c(1P)$ doublet, while the interpretations of other resonances are in dispute. For the resonances lying in the range of $2900\sim 3000~\rm{MeV}$, there exist two resonances $\Xi_c(2930)$ and $\Xi_c(2970)$, which were reported by the BaBar and Belle Collaborations, respectively~\cite{Aubert:2007eb,Chistov:2006zj}. Theoretical interpretations on $\Xi_c(2930)$ and $\Xi_c(2970)$ resonances include conventional charmed strange baryons~\cite{Ebert:2011kk,Chen:2014nyo,Chen:2016iyi,Chen:2007xf,Liu:2012sj,Mu:2014iaa,Ye:2017yvl,Ye:2017dra,Wang:2017kfr,Valcarce:2008dr,Chen:2015kpa,Cheng:2006dk,
Cheng:2015naa,Cheng:2017ove,Aliev:2018ube,Shah:2016nxi,Jia:2019bkr} and molecular states~\cite{JimenezTejero:2009vq,Lu:2014ina,Yu:2018yxl,Nieves:2019jhp} with various quantum numbers. The observations of LHCb Collaboration indicate that the $\Xi_c(2930)^0$ should be the overlap of the two narrow states $\Xi_c(2923)^0$ and $\Xi_c(2939)^0$, and whether the $\Xi_c(2970)^0$ and $\Xi_c(2965)^0$ structures are different or not needs further investigations~\cite{Aaij:2020yyt}. Above 3000 MeV, the situation becomes more complicated, and the detailed explanations and discussions can be found in the reviews~\cite{Chen:2016spr,Cheng:2015iom,Crede:2013sze}. It can been seen that the low-lying $\Xi_c$ and $\Xi_c^\prime$ spectra are far from being established.

The observations of $\Xi_c(2923)^0$, $\Xi_c(2939)^0$ and $\Xi_c(2965)^0$ resonances provide a good opportunity to study the low-lying $\Xi_c$ and $\Xi_c^\prime$ spectra. Compared with the predictions of constituent quark models~\cite{Ebert:2007nw,Roberts:2007ni,Ebert:2011kk,Chen:2014nyo,Chen:2016iyi}, these three resonances lie in the mass region of $\lambda-$mode $\Xi_c(2S)$, $\Xi_c^\prime(2S)$, $\Xi_c^{\prime *}(2S)$, and $\Xi_c^\prime(1P)$ states. Although the strong decay behaviors of these low-lying $\Xi_c$ and $\Xi_c^\prime$ states have been studied by several works within the quark models~\cite{Chen:2014nyo,Chen:2016iyi,Chen:2007xf,Liu:2012sj,Mu:2014iaa,Ye:2017yvl,Ye:2017dra,Wang:2017kfr}, it is essential to clarify the newly observed $\Xi_c(2923)^0$, $\Xi_c(2939)^0$ and $\Xi_c(2965)^0$ resonances into the $\Xi_c$ or $\Xi_c^\prime$ family. Due to the lack of the accurate experimental information, the previous works did not agree with each other and can hardly establish the low-lying $\Xi_c$ and $\Xi_c^\prime$ spectra. In fact, the $\Xi_c(2930)^0$ which is absent in present experimental observation, may have caused lots of troubles in previous studies.

In this issue, we calculate the strong decay behaviors of the newly observed $\Xi_c(2923)^0$, $\Xi_c(2939)^0$ and $\Xi_c(2965)^0$ resonances under various assignments within the $^3P_0$ model. Our results suggest that the $\Xi_c(2923)^0$ and $\Xi_c(2939)^0$ can be clarified into the $\lambda-$mode $J^P=3/2^-$ and $5/2^-$ $\Xi_c^\prime(1P)$ states respectively, and the $\Xi_c(2965)^0$ can be assigned as the $\lambda-$mode $J^P=1/2^+$ $\Xi_c^\prime(2S)$ state. Meanwhile, the strong decays of some missing partners are also presented, which may provide valuable information for future experimental searches.

This paper is organized as follows. The $^3P_0$ model and notations are introduced in Sec.~\ref{method}. The strong decay behaviors of the $\Xi_c(2923)^0$, $\Xi_c(2939)^0$ and $\Xi_c(2965)^0$ resonances are estimated and discussed in Sec.~\ref{decay}. A summary is given in the last section.

\section{$^3P_0$ model and notations}{\label{method}}

In present work, the $^3P_0$ model is adopted to estimate the strong decays of the $\Xi_c(2923)^0$, $\Xi_c(2939)^0$ and $\Xi_c(2965)^0$ resonances. This model has been extensively used for the strong decay behaviors of conventional hadrons and made great successes. There exist plenty of literatures on the $^3P_0$ model and some details can be found in Refs.~\cite{Chen:2016iyi,Chen:2007xf,Ye:2017yvl,micu,3p0model1,3p0model2,Lu:2018utx,Liang:2019aag}. Here, we only present the main ingredients of the $^3P_0$ model. For a $\Xi_c$ baryon, the transition operator $T$ of the decay $A\rightarrow BC$ is given by
\begin{eqnarray}
T&=&-3\gamma\sum_m\langle 1m1-m|00\rangle\int
d^3\boldsymbol{p}_4d^3\boldsymbol{p}_5\delta^3(\boldsymbol{p}_4+\boldsymbol{p}_5)\nonumber\\
&&\times {\cal{Y}}^m_1\left(\frac{\boldsymbol{p}_4-\boldsymbol{p}_5}{2}\right
)\chi^{45}_{1,-m}\phi^{45}_0\omega^{45}_0b^\dagger_{4i}(\boldsymbol{p}_4)d^\dagger_{4j}(\boldsymbol{p}_5),
\end{eqnarray}
where $\gamma$ is a dimensionless $q_4\bar{q}_5$ pair creation strength, and $\boldsymbol{p}_4$ and $\boldsymbol{p}_5$ are the momenta of the created quark $q_4$ and antiquark $\bar{q}_5$, respectively. The solid harmonic polynomial ${\cal{Y}}^m_1(\boldsymbol{p})\equiv|p|Y^m_1(\theta_p, \phi_p)$ reflects the $P-$wave distribution of the $q_4\bar{q}_5$ in the momentum space. $\phi^{45}_{0}=(u\bar u + d\bar d +s\bar s)/\sqrt{3}$, $\omega^{45}=\delta_{ij}$, and $\chi_{{1,-m}}^{45}$ are the flavor, color, and spin wave functions of the $q_4\bar{q}_5$, respectively.

With the transition operator, the helicity amplitude ${\cal{M}}^{M_{J_A}M_{J_B}M_{J_C}}$ is defined as
\begin{eqnarray}
\langle
BC|T|A\rangle=\delta^3(\boldsymbol{P}_A-\boldsymbol{P}_B-\boldsymbol{P}_C){\cal{M}}^{M_{J_A}M_{J_B}M_{J_C}}.
\end{eqnarray}
The explicit formula of the helicity amplitude can be found in Refs.~\cite{Chen:2016iyi,Chen:2007xf,Ye:2017yvl,Lu:2018utx,Liang:2019aag}. Then, the decay width of $A\to BC$ process can be obtained straightforward
\begin{eqnarray}
\Gamma= \pi^2\frac{p}{M^2_A}\frac{1}{2J_A+1}\sum_{M_{J_A},M_{J_B},M_{J_C}}|{\cal{M}}^{M_{J_A}M_{J_B}M_{J_C}}|^2,
\end{eqnarray}
where $p=|\boldsymbol{p}|$ is the momentum of the final hadrons in the center of mass system.

The notations of relevant initial states and the predicted masses from quark models are listed in Table~\ref{tab1}. Here, the $\rho-$mode quantum numbers $n_\rho=l_\rho=0$ are omitted, since only the $\lambda-$mode $\Xi_c(2S)$, $\Xi_c^\prime(2S)$, $\Xi_c^{\prime *}(2S)$, and $\Xi_c^\prime(1P)$ states are considered. For the masses of these initial states, we first adopt the experimental values of $\Xi_c(2923)^0$, $\Xi_c(2939)^0$ and $\Xi_c(2965)^0$ resonances by assuming that they are possible candidates. If a assignment is finally disfavored, the predicted mass of this state is applied to recalculate its strong decays. The masses of final states are taken from the Review of Particle Physics~\cite{Tanabashi:2018oca}.

All the parameters in the $^3P_0$ model used here are the same as our previous works~\cite{Lu:2018utx,Liang:2019aag,Lu:2019rtg,Liang:2020hbo}, which have been employed to describe the strong decay behaviors of various singly heavy baryons successfully. More specifically, the effective value $R= 2.5~\rm{GeV^{-1}}$ is adopted for the pseudoscalar mesons~\cite{Godfrey:2015dva}, while the $\alpha_\rho=400~\rm{MeV}$, $420~\rm{MeV}$, and $440~\rm{MeV}$ are applied for the $\Lambda_{c(b)}$ and $\Sigma_{c(b)}$, $\Xi_{c(b)}$ and $\Xi^\prime_{c(b)}$, and $\Omega_{c(b)}$, respectively~\cite{Lu:2018utx,Zhong:2007gp}. The $\alpha_\lambda$ can be obtained
\begin{eqnarray}
\alpha_\lambda=\Bigg(\frac{3m_Q}{m_{q_1}+m_{q_2}+m_Q} \Bigg)^{1/4} \alpha_\rho,
\end{eqnarray}
where the $m_Q$ and $m_{q_1}(m_{q_2})$ are the heavy and light quark masses, respectively. The $m_{u/d}=220~\rm{MeV}$, $m_s=419~\rm{MeV}$, and $m_c=1628~\rm{MeV}$ are introduced to explicitly take into account the quark mass differences~\cite{Godfrey:1985xj,Capstick:1986bm,Godfrey:2015dva}.
The overall parameter $\gamma$ equals to 9.83 is obtained by reproducing the well established process, and more discussions on this parameter can be found in Ref.~\cite{Liang:2020hbo}. The dependence of the harmonic parameter $\alpha_\rho$ for baryons will be discussed in the following section.

\begin{table}[!htbp]
\begin{center}
\caption{ \label{tab1} Notations, quantum numbers, and the predicted masses of the relevant states. The $n_\lambda$ and $l_\lambda$ are the nodal quantum number and orbital angular momentum between the light quark subsystem and the charm quark, respectively. $L$, $S_\rho$, and $j$ stand for the total orbital angular momentum, total spin of the two light quarks, total angular momentum of $L$ and $S_\rho$, respectively. $J^P$ represent the spin-party of the hadron. The units are in MeV.}
\renewcommand{\arraystretch}{1.5}
\footnotesize
\begin{tabular*}{8.6cm}{@{\extracolsep{\fill}}*{9}{p{0.8cm}<{\centering}}}
\hline\hline
 State                              &  $n_\lambda$      &  $l_\lambda$  &  $L$  & $S_\rho$ & $j$  & $J^P$         & RM~\cite{Ebert:2011kk}  & NR~\cite{Chen:2016iyi}       \\\hline
 $\Xi_c(2S)$                        &  1                &  0            &  0    & 0        & 0      & $\frac{1}{2}^+$    & 2959  & 2940 \\
 $\Xi_c^\prime(2S)$                 &  1                &  0            &  0    & 1        & 1      & $\frac{1}{2}^+$   & 2983   & 2977 \\
 $\Xi_c^{\prime *}(2S)$              &  1                &  0            &  0    & 1        & 1      & $\frac{3}{2}^+$  & 3026    & 3007 \\
 $\Xi^\prime_{c0}(\frac{1}{2}^-)$   &  0                &  1            &  1    & 1        & 0      & $\frac{1}{2}^-$   & 2936   & 2839 \\
 $\Xi^\prime_{c1}(\frac{1}{2}^-)$   &  0                &  1            &  1    & 1        & 1      & $\frac{1}{2}^-$   & 2854   & 2900 \\
 $\Xi^\prime_{c1}(\frac{3}{2}^-)$   &  0                &  1            &  1    & 1        & 1      & $\frac{3}{2}^-$  & 2935    & 2932 \\
 $\Xi^\prime_{c2}(\frac{3}{2}^-)$   &  0                &  1            &  1    & 1        & 2      & $\frac{3}{2}^-$  & 2912    & 2921 \\
 $\Xi^\prime_{c2}(\frac{5}{2}^-)$   &  0                &  1            &  1    & 1        & 2      & $\frac{5}{2}^-$   & 2929   & 2927 \\
 \hline\hline
\end{tabular*}
\end{center}
\end{table}

\section{Strong decays}{\label{decay}}

\subsection{$\Xi_c(2S)$ state}

In the constituent quark model, only one $\lambda-$mode $\Xi_c(2S)$ state exists. From Table~\ref{tab1}, the predicted masses of the $\Xi_c(2S)$ state are 2959 and 2940 MeV within the relativistic and nonrelativistic quark models, respectively. The strong decay behaviors of the $\Xi_c(2923)^0$, $\Xi_c(2939)^0$ and $\Xi_c(2965)^0$ under $\Xi_c(2S)$ assignments are presented in Table~\ref{xi2s1}. It is shown that the total decay widths for the $\Xi_c(2923)^0$, $\Xi_c(2939)^0$ and $\Xi_c(2965)^0$ are predicted to be 4.41, 5.31, and 7.66 MeV, respectively. The calculated total decay widths are smaller than the experimental data. Moreover, the $\Lambda_c \bar K$ decay mode is forbidden for the $\Xi_c(2S)$ state due to the quantum number conservation, which contradicts with the experimental observations. Thus, the $\Xi_c(2S)$ assignment can be totally excluded.

\begin{table}[!htbp]
\begin{center}
\caption{\label{xi2s1} Strong decays of the $\Xi_c(2923)^0$, $\Xi_c(2939)^0$ and $\Xi_c(2965)^0$ under $\Xi_c(2S)$ assignments in MeV.}
\renewcommand{\arraystretch}{1.5}
\begin{tabular*}{8.6cm}{@{\extracolsep{\fill}}*{4}{p{2cm}<{\centering}}}
\hline\hline
	&	\multicolumn{3}{c}{$\Xi_c(2S)$}    \\\cline{2-4}
Mode	&	$\Xi_c(2923)^0$       &  $\Xi_c(2939)^0$  &  $\Xi_c(2965)^0$  	\\\hline
$\Xi_c^\prime \pi$	    &	2.36	     & 2.74     &  3.44  \\
$\Xi_c^{\prime *} \pi$	&	2.05	     & 2.57     &  3.58  \\
$\Sigma_c \bar K$	    &	$\cdot \cdot \cdot$	    & $\cdot \cdot \cdot$     &  0.64  \\
Total width	            &	4.41         & 5.31     &  7.66  \\
Experiments 	&	$7.1\pm0.8\pm1.8$ & $10.2\pm0.8\pm1.1$ & $14.1\pm0.9\pm1.3$ \\
\hline\hline
\end{tabular*}
\end{center}
\end{table}

Given the flavor symmetry, the mass gap between $\Xi_c(2S)$ and $\Xi_c$ should be similar with the non-strange $\Lambda_c$ case, that is
\begin{eqnarray}
m[\Xi_c(2S)]- m[\Xi_c]  \simeq  m[\Lambda_c(2S)]- m[\Lambda_c] = 480~\rm{MeV}.
\end{eqnarray}
With this approximate equality, the mass of $\Xi_c(2S)$ state should be around 2951 MeV, which agrees well with the quark model predictions. We adopt the 2959 MeV predicted by relativistic quark model to calculate the strong decays for the $\Xi_c(2S)$. From Table~\ref{xi2s2}, it can be seen that the $\Xi_c(2S)$ state should be a rather narrow state, and the branching ratios for this state are predicted to be
\begin{eqnarray}
Br(\Xi_c^\prime \pi, \Xi_c^{\prime *} \pi, \Sigma_c \bar K) = 47.3\%, 48.2\%, 4.5\%,
\end{eqnarray}
which is independent with the overall strength $\gamma$. The predicted narrow width of $\Xi_c(2S)$ state here is consistent with that of the $^3P_0$ model~\cite{Chen:2007xf,Ye:2017dra} and chiral quark model~\cite{Liu:2012sj}, but quite different with the result of potential model~\cite{Chen:2016iyi}. The predicted branching ratios of ours are consistent with these works~\cite{Ye:2017dra,Liu:2012sj,Chen:2016iyi}, which suggests that the future experiments can search for the $\Xi_c(2S)$ state in the $\Xi_c^\prime \pi$ and $\Xi_c^{\prime *} \pi$ final states.

\begin{table}[!htbp]
\begin{center}
\caption{\label{xi2s2} Strong decays of the $\Xi_c(2S)$ state with a mass of 2959 MeV.}
\renewcommand{\arraystretch}{1.5}
\begin{tabular*}{8.6cm}{@{\extracolsep{\fill}}*{2}{p{2cm}<{\centering}}}
\hline\hline
Mode	&	$\Xi_c(2S)$ 	\\\hline
$\Xi_c^\prime \pi$	       &  3.28  \\
$\Xi_c^{\prime *} \pi$	  &  3.34  \\
$\Sigma_c \bar K$	       &  0.31  \\
Total width	            &  6.93  \\
\hline\hline
\end{tabular*}
\end{center}
\end{table}

\subsection{$\Xi_c^\prime(2S)$ and $\Xi_c^{\prime *}(2S)$ states}

The strong decays of the $\Xi_c(2923)^0$, $\Xi_c(2939)^0$ and $\Xi_c(2965)^0$ as $\Xi_c^\prime(2S)$ and $\Xi_c^{\prime *}(2S)$ states are calculated and listed in Table~\ref{xicp2s1}. It is shown that the predicted widths of these assignments are somewhat larger than the experimental data. Given the uncertainties of the $^3P_0$ model, these assignments seem to be acceptable. However, from Table~\ref{tab1}, the predicted masses of the $\Xi_c^\prime(2S)$ and $\Xi_c^{\prime *}(2S)$ are around 2980 and 3010 MeV, which are significantly larger than the $\Xi_c(2923)^0$ and $\Xi_c(2939)^0$ resonances. When the masses and decay widths are considered together, the $\Xi_c(2965)^0$ as $\Xi_c^\prime(2S)$ state is favored and other assignments are disfavored.
\begin{table*}
\begin{center}
\caption{\label{xicp2s1} Strong decays of the $\Xi_c(2923)^0$, $\Xi_c(2939)^0$ and $\Xi_c(2965)^0$ as $\Xi_c^\prime(2S)$ and $\Xi_c^{\prime *}(2S)$ states in MeV.}
\renewcommand{\arraystretch}{1.5}
\begin{tabular*}{18cm}{@{\extracolsep{\fill}}*{7}{p{2cm}<{\centering}}}
\hline\hline
	    &	\multicolumn{3}{c}{$\Xi_c^\prime(2S)$} & \multicolumn{3}{c}{$\Xi_c^{\prime *}(2S)$}  \\ \cline{2-4} \cline{5-7}
Mode	&	$\Xi_c(2923)^0$       &  $\Xi_c(2939)^0$  &  $\Xi_c(2965)^0$   &	$\Xi_c(2923)^0$       &  $\Xi_c(2939)^0$  &  $\Xi_c(2965)^0$	\\\hline
$\Xi_c \pi$	            &	5.36	     & 5.84     &  6.64                & 5.36            &   5.84    & 6.64\\
$\Xi_c^\prime \pi$	    &	3.14	     & 3.65     &  4.58                & 0.79            &   0.91    & 1.15 \\
$\Xi_c^{\prime *} \pi$	&	0.69	     & 0.86     &  1.20                & 1.71            &   2.14    & 2.99 \\
$\Lambda_c \bar K$	    &	4.76	     & 5.35     &  6.29                & 4.76            &   5.35    & 6.29 \\
$\Sigma_c \bar K$	    &$\cdot \cdot \cdot$ & $\cdot \cdot \cdot$ &  0.86 & $\cdot \cdot \cdot$ & $\cdot \cdot \cdot$    & 0.21                \\
Total width	            &   13.95         & 15.70     & 19.57                 & 12.62            &   14.24    & 17.28 \\
Experiments 	&	$7.1\pm0.8\pm1.8$ & $10.2\pm0.8\pm1.1$ & $14.1\pm0.9\pm1.3$   &	$7.1\pm0.8\pm1.8$ & $10.2\pm0.8\pm1.1$ & $14.1\pm0.9\pm1.3$ \\
\hline\hline
\end{tabular*}
\end{center}
\end{table*}

\begin{figure}[htb]
\includegraphics[scale=0.7]{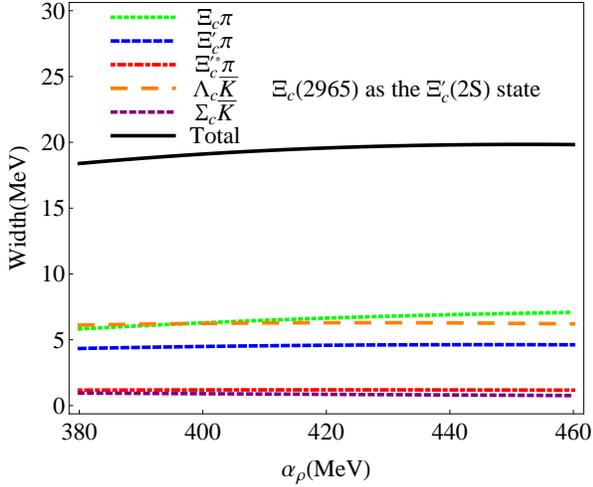}
\vspace{0.0cm} \caption{The dependence on the harmonic oscillator parameter $\alpha_\rho$ of the $\Xi_c(2965)^0$ as $\Xi_c^\prime(2S)$ state. When the $\alpha_\rho$ for the $\Xi_c(2965)^0$ varies from 380 to 460 MeV, the corresponding ones for the final non-strange states change in the range of $360 \sim 440~\rm{MeV}$.}
\label{xic2s}
\end{figure}

Two factors, the harmonic oscillator parameter $\alpha_\rho$ and overall strength $\gamma$, may affect the final total decay width. We plot the decay widths of the $\Xi_c(2965)^0$ under $\Xi_c^\prime(2S)$ assignment versus the $\alpha_\rho$ in Fig.~\ref{xic2s}.  It can be seen that when the $\alpha_\rho$ varies, the partial and total decay widths are almost unchanged. Within the reasonable range of $\alpha_\rho$, our conclusions remain. The uncertainties arising from the overall constant $\gamma$ can be eliminated when the branching ratios are concentrated. The predicted branching ratios of dominating channels for the $\Xi_c(2965)^0$ are
\begin{eqnarray}
Br(\Xi_c \pi, \Xi_c^\prime \pi, \Lambda_c \bar K) = 34.0\%, 23.4\%, 32.2\%,
\end{eqnarray}
which is also consistent with experimental observation in the $\Lambda_c \bar K$ mass spectrum.

As mentioned in the Introduction, the $\Xi_c(2965)^0$ may be the corresponding charmed strange partner of the $\Omega_c(3090)$. Although various interpretations of the $\Omega_c(3090)$ exist, the predicted mass of $\Omega_c(2S)$ state is 3088 MeV in the relativistic quark model~\cite{Ebert:2011kk}, which indicates that the $\Omega_c(3090)$ as the $\Omega_c(2S)$ state is possible. Moreover, a structure $\Sigma_c(2850)^0$ with $2846~\rm{MeV}$ have been reported by the BaBar Collaboration~\cite{Aubert:2008ax}, and the mass gap between $\Xi_c(2965)^0$ and $\Sigma_c(2850)^0$ is
\begin{eqnarray}
m[\Xi_c(2965)^0] - m[\Sigma_c(2850)^0] = 119~\rm{MeV}.
\end{eqnarray}
This mass gap is similar with $m[\Omega_c(3090)^0] - m[\Xi_c(2965)^0]$, which suggests that the $\Sigma_c(2850)^0$ structure may be the nonstrange partner of $\Xi_c(2965)^0$. Meanwhile, the mass and strong decay behaviors of $\Sigma_c(2850)^0$ suggests that it should correspond to the $\Sigma_c(2S)$ state~\cite{Chen:2016iyi,Cheng:2017ove}. All these evidences support $\Xi_c(2965)^0$ as the $\Xi_c^\prime(2S)$ state. Further studies on the low-lying $\Sigma_c$, $\Xi_c^\prime$ and $\Omega_c$ states may reveal more connections among these three families.

When we regard the $\Xi_c(2965)^0$ as $\Xi_c^\prime(2S)$ state, the mass of $\Xi_c^{\prime *}(2S)$ state can be estimated via the fine splitting. In the traditional quark model, the fine splitting of $2S$ states should be smaller than the corresponding $1S$ ones, that is
\begin{eqnarray}
m[\Xi_c^{\prime *}(2S)] - m[\Xi_c^\prime(2S)] <  m[\Xi_c^{\prime *}] - m[\Xi_c^\prime] = 67~\rm{MeV},
\end{eqnarray}
which is consistent with the predicted fine splitting by the quark models~\cite{Ebert:2011kk,Chen:2016iyi}. Here, we adopt the 3007 MeV to calculate the strong decays of $\Xi_c^{\prime *}(2S)$. From Table~\ref{xicp2s2}, the predicted total decay width is about 23 MeV, and the dominating decay modes are $\Xi_c \pi$, $\Xi_c^{\prime *} \pi$, and $\Lambda_c \bar K$. The branching ratios are
\begin{eqnarray}
Br(\Xi_c \pi, \Xi_c^{\prime *} \pi, \Lambda_c \bar K) = 34.1\%, 20.0\%, 32.7\%,
\end{eqnarray}
which are independent with the quark pair creation strength $\gamma$ and can be tested by future experiments.
\begin{table}
\begin{center}
\caption{\label{xicp2s2} Strong decays of the $\Xi_c^{\prime *}(2S)$ state with a mass of 3007 MeV.}
\renewcommand{\arraystretch}{1.5}
\begin{tabular*}{8.6cm}{@{\extracolsep{\fill}}*{2}{p{2cm}<{\centering}}}
\hline\hline
Mode	    & $\Xi_c^{\prime *}(2S)$  \\ \hline	
$\Xi_c \pi$	            &	7.89	    \\
$\Xi_c^\prime \pi$	    &	1.55	      \\
$\Xi_c^{\prime *} \pi$	&	4.63	     \\
$\Lambda_c \bar K$	    &	7.57	    \\
$\Sigma_c \bar K$	    &   1.48  \\
Total width	            &   23.12       \\
\hline\hline
\end{tabular*}
\end{center}
\end{table}

\subsection{$\Xi_c^\prime(1P)$ states}

Five $\lambda-$mode $\Xi_c^\prime(1P)$ states, denoted as $\Xi^\prime_{c0}(\frac{1}{2}^-)$, $\Xi^\prime_{c1}(\frac{1}{2}^-)$, $\Xi^\prime_{c1}(\frac{3}{2}^-)$, $\Xi^\prime_{c2}(\frac{3}{2}^-)$, and $\Xi^\prime_{c2}(\frac{5}{2}^-)$, are allowed in the conventional quark model. From Table~\ref{tab1}, it is shown that the predicted masses are in the range of $2854 \sim 2936$ MeV, which indicates that the $\Xi_c(2923)^0$ and $\Xi_c(2939)^0$ are good candidates of these $\Xi_c^\prime(1P)$ states. Although the $\Xi_c(2965)^0$ lies higher than the predicted masses and can be assigned as the $\Xi_c^{\prime}(2S)$ state, the possibility of $\Xi_c(2965)^0$ as $\Xi_c^\prime(1P)$ states are also calculated. The total decay widths with various assignments are presented in Table~\ref{xicp}. For the $j = 0$ state, the predicted total decay width is rather large, which can be fully excluded. For the two $j=1$ states, the total decay widths also seem larger than the experimental data and the $\Lambda_c \bar K$ decay mode is forbidden due to the quantum number conservation. For the two $j = 2$ states, the calculated total decay widths agree well with the experimental data, which indicate that all the $\Xi_c(2923)^0$, $\Xi_c(2939)^0$ and $\Xi_c(2965)^0$ structures can be regarded as the $\Xi^\prime_{c2}(\frac{3}{2}^-)$ and $\Xi^\prime_{c2}(\frac{5}{2}^-)$ states. To describe these three resonances simultaneously, we prefer the normal mass order and assign the $\Xi_c(2923)^0$, $\Xi_c(2939)^0$ and $\Xi_c(2965)^0$ as $\Xi^\prime_{c2}(\frac{3}{2}^-)$, $\Xi^\prime_{c2}(\frac{5}{2}^-)$ and $\Xi_c^{\prime}(2S)$ states, respectively.

\begin{table}[!htbp]
\begin{center}
\caption{\label{xicp} Total decay widths of the $\Xi_c(2923)^0$, $\Xi_c(2939)^0$ and $\Xi_c(2965)^0$ under various $\Xi_c^\prime(1P)$ assignments in MeV.}
\renewcommand{\arraystretch}{1.5}
\begin{tabular*}{8.6cm}{@{\extracolsep{\fill}}*{4}{p{2cm}<{\centering}}}
\hline\hline
	State                    &	$\Xi_c(2923)^0$       &  $\Xi_c(2939)^0$  &  $\Xi_c(2965)^0$  	\\\hline
$\Xi^\prime_{c0}(\frac{1}{2}^-)$	    &	306.83	     & 304.83     &  297.21  \\
$\Xi^\prime_{c1}(\frac{1}{2}^-)$	    &	94.23	     & 98.22     &  249.94  \\
$\Xi^\prime_{c1}(\frac{3}{2}^-)$	    &	72.39	     & 78.42     &  87.88  \\
$\Xi^\prime_{c2}(\frac{3}{2}^-)$	    &	8.56         & 10.66      &  15.01  \\
$\Xi^\prime_{c2}(\frac{5}{2}^-)$	    &	8.24         & 10.25      &  14.42  \\
Experiments 	             &	$7.1\pm0.8\pm1.8$ & $10.2\pm0.8\pm1.1$ & $14.1\pm0.9\pm1.3$ \\
\hline\hline
\end{tabular*}
\end{center}
\end{table}

\begin{figure}[htb]
\includegraphics[scale=0.27]{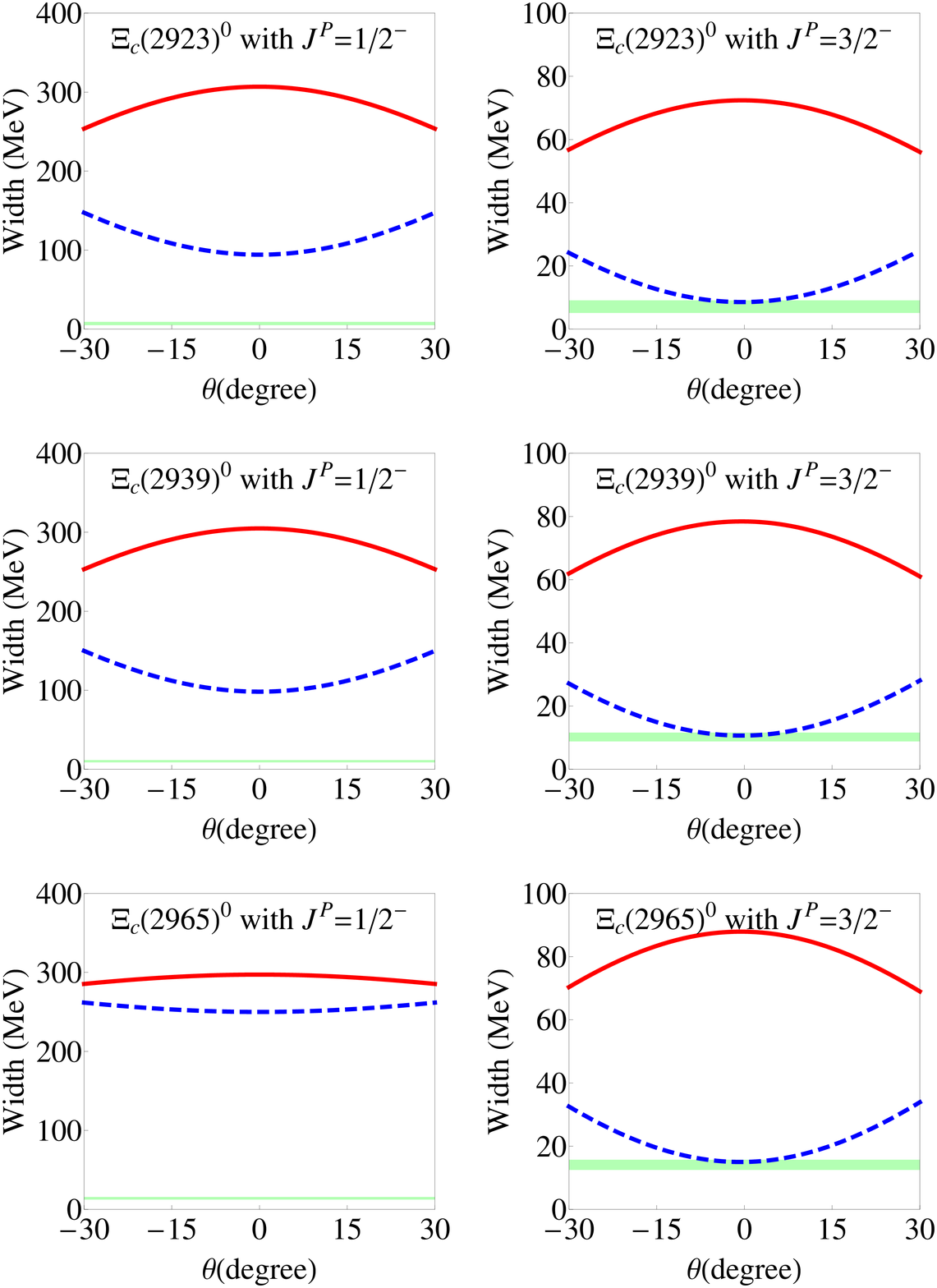}
\vspace{0.0cm} \caption{The total decay widths under various assignments as functions of the mixing angle $\theta$ in the range of $-30^\circ \sim 30^\circ$. The red solid lines stand for the $|1P~1/2^- \rangle_1$ and $|1P~3/2^- \rangle_1$ states, and the blue dashed curves
are the $|1P~1/2^- \rangle_2$ and $|1P~3/2^- \rangle_2$ states. The green bands
stand for the experimental total decay widths.}
\label{xic1p}
\end{figure}

Moreover, the physical structures can be the mixing of the quark model states with same $J^P$, that is to say
\begin{equation}
\left(\begin{array}{c}| 1 P~{1/2^-}\rangle_1\cr | 1 P~{1/2^-}\rangle_2
\end{array}\right)=\left(\begin{array}{cc} \cos\theta & \sin\theta \cr -\sin\theta &\cos\theta
\end{array}\right)
\left(\begin{array}{c} |1/2^-,j=0
\rangle \cr |1/2^-,j=1\rangle
\end{array}\right),
\end{equation}
\begin{equation}
\left(\begin{array}{c}|1 P~{3/2^-}\rangle_1\cr | 1 P~{3/2^-}\rangle_2
\end{array}\right)=\left(\begin{array}{cc} \cos\theta & \sin\theta \cr -\sin\theta &\cos\theta
\end{array}\right)
\left(\begin{array}{c} |3/2^-,j=1
\rangle \cr |3/2^-,j=2\rangle
\end{array}\right).
\end{equation}
In the heavy quark limit, these mixing angles should be zero and the heavy quark symmetry is preserved. The finite charm quark mass may break this symmetry explicitly, and the physical states and quark model states may have a small divergence with a nonzero mixing angle. The dependence on the mixing angle $\theta$ in the range of $-30^\circ \sim 30^\circ$ are presented in Figure~\ref{xic1p}. It is shown that the two $J^P=1/2^-$ states can be excluded, while the $J^P=3/2^-$ assignments are allowed. From the mixing scheme, all the $\Xi_c(2923)^0$, $\Xi_c(2939)^0$ and $\Xi_c(2965)^0$ may belong to the narrower $|1P~3/2^- \rangle_2$ state. Actually, compared with the experimental data, the mixing angle should be extremely small, and the $|1P~3/2^- \rangle_2$ state is almost same as the pure $\Xi^\prime_{c2}(\frac{3}{2}^-)$ state. Thus, the mixing effects can be neglected here, and our above assignments of these three resonances remain. These interpretations on the $\Xi_c(2923)^0$, $\Xi_c(2939)^0$ and $\Xi_c(2965)^0$ are different with the results within QCD sum rules, where all the resonances are clarified into the $\Xi_c^\prime(1P)$ states~\cite{Yang:2020zjl}.

The partial decay widths for the $\Xi_c(2923)^0$ and $\Xi_c(2939)^0$ under $j=2$ assignments are also presented in Table~\ref{xicp2} for reference. It is shown that the $\Xi_c \pi$, and $\Lambda_c \bar K$ are the dominating decay modes for these two states. Under our present interpretations, three $P-$wave states, $\Xi^\prime_{c0}(\frac{1}{2}^-)$, $\Xi^\prime_{c1}(\frac{1}{2}^-)$ and $\Xi^\prime_{c1}(\frac{3}{2}^-)$, have not been observed. For the $j=0$ state, it may be hardly found due to its larger total decay width. For the two $j=1$ states, the $\Xi_c \pi$ and $\Lambda_c \bar K$ channels are forbidden, and other final states, such as $\Xi_c^\prime \pi$ and $\Xi_c^{\prime *} \pi$, can help us to hunt for them. Moreover, the similarities among $\Sigma_c$, $\Xi_c^\prime$ and $\Omega_c$ spectra may also provide valuable clues, and more experimental information and theoretical efforts on these flavor $6_F$ states are needed.
\begin{table}
\begin{center}
\caption{\label{xicp2} Strong decays of the $\Xi_c(2923)^0$ and $\Xi_c(2939)^0$ under $j=2$ assignments in MeV.}
\renewcommand{\arraystretch}{1.5}
\begin{tabular*}{8.6cm}{@{\extracolsep{\fill}}*{3}{p{2.8cm}<{\centering}}}
\hline\hline
Mode	&	$\Xi_c(2923)^0$ as $\Xi^\prime_{c2}(\frac{3}{2}^-)$  &  $\Xi_c(2939)^0$ as $\Xi^\prime_{c2}(\frac{5}{2}^-)$ \\\hline
$\Xi_c \pi$	            &	4.79	     & 5.72           \\
$\Xi_c^\prime \pi$	    &	0.76	     & 0.44             \\
$\Xi_c^{\prime *} \pi$	&	0.18	     & 0.42            \\
$\Lambda_c \bar K$	    &	2.83	     & 3.67          \\
Total width	            &   8.56         & 10.25          \\
Experiments 	&	$7.1\pm0.8\pm1.8$ & $10.2\pm0.8\pm1.1$  \\
\hline\hline
\end{tabular*}
\end{center}
\end{table}

In the Review of Particle Physics~\cite{Tanabashi:2018oca}, there also exist two states, $\Xi_c(2930)$ and $\Xi_c(2970)$.  The $\Xi_c(2930)^+$ and $\Xi_c(2930)^0$ have total widths of 15 and 26 MeV, respectively, which can not be clarified into our present arrangement. We agree with the suggestion of LHCb Collaboration that the $\Xi_c(2930)$ is the overlap of two structures $\Xi_c(2923)$ and $\Xi_c(2939)$~\cite{Aaij:2020yyt}. The $\Xi_c(2970)$ have a nearly identical mass with $\Xi_c(2965)$, and the measured widths by different collaborations show large divergence. Theoretically, it has been interpreted as the $\Xi_c(2S)$, $\Xi_c^\prime(2S)$, and $\Xi_c^\prime(1P)$ states. Based on our calculations, the predicted total decay width of $\Xi_c(2S)$ state is rather small, which does not support the $\Xi_c(2970)$ as $\Xi_c(2S)$ state. Moreover, the mass gap between $\Xi_c(2970)$ and $\Xi_c(2965)$ contradicts with the mass splitting of $\Xi_c(2S)$ and $\Xi_c^\prime(2S)$ states. In the $P-$wave states, there is also no room left for the $\Xi_c(2970)$ under our assignments. Together with the mass and total width, we suggest that the $\Xi_c(2970)$ should be the same state as $\Xi_c(2965)$. More information on spin-parity and branching ratios can help us to clarify its nature.

\section{Summary}{\label{Summary}}

 In this work, we estimate the strong decays of three newly observed resonances $\Xi_c(2923)^0$, $\Xi_c(2939)^0$ and $\Xi_c(2965)^0$ by the LHCb Collaboration. Given the $\Xi_c$ and $\Xi_c^\prime $ spectra predicted by constituent quark models, these three resonances can be tentatively treated as the $\lambda-$mode $\Xi_c(2S)$, $\Xi_c^\prime(2S)$, $\Xi_c^{\prime *}(2S)$, and $\Xi_c^\prime(1P)$ states. Their strong decay behaviors are calculated within the $^3P_0$ model. Compared with the experimental data, our results indicate that the $\Xi_c(2923)^0$ and $\Xi_c(2939)^0$ should be $J^P=3/2^-$ and $5/2^-$ $\Xi_c^\prime(1P)$ states respectively, and the $\Xi_c(2965)^0$ can be assigned as the $J^P=1/2^+$ $\Xi_c^\prime(2S)$ state. Also, we suggest that the previous observed $\Xi_c(2930)$ may be the overlap of two structures $\Xi_c(2923)$ and $\Xi_c(2939)$, and the $\Xi_c(2970)$ should be the same state as $\Xi_c(2965)$. Other theoretical results of the missing $\Xi_c(2S)$, $\Xi_c^{\prime *}(2S)$, and $\Xi_c^\prime(1P)$ states may be helpful for future experiments.

 During the study, it can be noticed that the $\Xi_c$ and $\Xi_c^\prime$ systems are more complicated than other singly heavy baryons because the flavor configurations can not be determined by the isospin quantum number. Fortunately, there exist some similarities and connections among the $\Sigma_c$, $\Xi_c^\prime$ and $\Omega_c$ baryons, which can provide valuable clues for us. More theoretical and experimental studies on these three families are needed to further understand their inner structures and establish the low-lying spectra.

\bigskip
\noindent
\begin{center}
{\bf ACKNOWLEDGEMENTS}\\

\end{center}
I would like to thank Xian-Hui Zhong, Hua-Xing Chen, Wei Liang, Li-Ye Xiao, Guang-Juan Wang, and Kai-Lei Wang for valuable discussions on the singly heavy baryons. This work is supported by the National Natural Science Foundation of China under Grants No.~11705056 and No.~U1832173.

\end{document}